\documentclass{article}

\usepackage{arxiv}

\usepackage[utf8]{inputenc} % allow utf-8 input
\usepackage[T1]{fontenc}    % use 8-bit T1 fonts
\usepackage{hyperref}       % hyperlinks
\usepackage{url}            % simple URL typesetting
\usepackage{booktabs}       % professional-quality tables
\usepackage{amsfonts}       % blackboard math symbols
\usepackage{amsmath}        % for align* environment
\usepackage{nicefrac}       % compact symbols for 1/2, etc.
\usepackage{microtype}      % microtypography
\usepackage{graphicx}
\usepackage[round]{natbib}
\usepackage{doi}
\usepackage{booktabs}
\usepackage{float}
\usepackage{threeparttable}

\title{Don't Measure Once: Measuring Visibility in AI Search (GEO)}

\author{\href{https://orcid.org/0009-0000-9890-403X}{\includegraphics[scale=0.06]{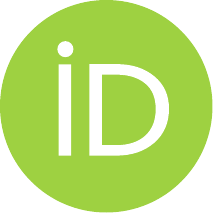}\hspace{1mm}Julius Schulte}\thanks{Affiliated with \href{https://aurora-intelligence.tech/}{Aurora Intelligence}.}\\
	Global Center for Entrepreneurship and Innovation\\
	University of St. Gallen\\
	9000 St.Gallen, Switzerland \\
	\texttt{julius.schulte@unisg.ch} \\
    \And
    \href{https://orcid.org/0009-0002-5638-6750}{\includegraphics[scale=0.06]{orcid.pdf}\hspace{1mm}Malte Bleeker}\\
	Marketing und Customer Insight\\
	University of St. Gallen\\
	9000 St.Gallen, Switzerland \\
	\texttt{malte.bleeker@unisg.ch} \\
	\And
    Philipp Kaufmann\\
	Marketing und Customer Insight\\
	University of St. Gallen\\
	9000 St.Gallen, Switzerland \\
	\texttt{philipp.kaufmann@unisg.ch} \\
}

\renewcommand{\headeright}{}
\renewcommand{\undertitle}{}
\hypersetup{
pdftitle={Don't Measure Once: Measuring Visibility in AI Search (GEO)},
pdfsubject={Generative Engine Optimization, AI Search, Information Retrieval},
pdfauthor={Julius Schulte, Philipp Kaufmann, Malte Bleeker},
pdfkeywords={AI Visibility, Generative Engine Optimization, GEO, SEO, Information Retrieval, LLM, Brand Visibility},
}

\begin{document}
\maketitle

\begin{abstract}
As large language model-based chat systems become increasingly widely used, generative engine optimization (GEO) has emerged as an important problem for information access and retrieval. In classical search engines, results are comparatively transparent and stable: a single query often provides a representative snapshot of where a page or brand appears relative to competitors. The inherent probabilistic nature of AI search changes this paradigm. Answers can vary across runs, prompts, and time, making one-off observations unreliable. Drawing on empirical studies, our findings underscore the need for repeated measurements to assess a brand's GEO performance and to characterize visibility as a distribution rather than a single-point outcome.
\end{abstract}

% keywords can be removed

\section{Introduction}

With the emergence of Large Language Models (LLMs), the way consumers retrieve information is undergoing a paradigm shift. This transformation poses a direct challenge to the traditional search ecosystem \citep{wenPositionRisksGenerative2025}, which remains the foundational marketing communication channel for most companies \citep{sturzeAgileMarketingPerformance2022}. Signs of a market restructuring are evident: Google, having historically monopolized search in the Western hemisphere, experienced its first market share drop in a decade in late 2024 \citep{goodwinGooglesSearchMarket2025}. In stark contrast, generative search (or LLM-based search) has seen explosive growth, exemplified by ChatGPT increasing its weekly active user base from $\approx$ 100 million (Jan. 2024) to ${\approx}$780 million (Sept. 2025) \citep{chatterjiHowPeopleUse2025a}. Some studies estimate that ChatGPT will surpass Google in four years \citep{AIVisibilitySEO}.
This trend reflects a broader change in user behavior, as consumers increasingly rely on AI agents not just for information retrieval, but for executing complex tasks such as shopping, planning, and coding \citep{economistHowAIDisrupting}.

This paradigm shift from traditional search to generative search fundamentally alters the mechanisms of performance monitoring. While digital marketers have historically benefited from high levels of data transparency in Search Engine Optimisation (SEO), which is facilitated by first-party utilities such as Google Search Console (GSC), the transition toward Generative Engine Optimisation (GEO) has introduced a significant observability gap \citep{aggarwalGEOGenerativeEngine2024a}. Unlike traditional search engines, the providers of LLMs do not currently offer native, proprietary monitoring tools equivalent to GSC. Consequently, foundational metrics such as which specific search queries are used and the corresponding query volumes are no longer directly observable within the GEO ecosystem.

In the absence of ground-truth data, marketers must adopt alternative methodologies to evaluate performance. The emerging industry standard is to measure visibility, which is defined as the frequency and prominence of brand mentions within generated responses \citep{rejon2025generative}. However, relying on static visibility alone is insufficient due to the stochastic nature of generative models. This study introduces stability as a critical, complementary performance dimension. Even as novel third-party tools are developed to quantify AI-specific visibility \citep{rejon2025generative}, marketers must avoid drawing premature conclusions from these snapshot metrics. Without accounting for the stability of mentions over time and across varying prompt iterations, visibility data remains prone to volatility and misinterpretation.

\section{Literature}

Recent empirical research provides behavioural evidence highlighting the shift from traditional search to generative search. Using click-stream data, \citet{padillaImpactLLMAdoption2025} show that adoption of AI-based search engines leads to a gradual but substantial decline in traditional search. Specifically, traditional search queries fell by more than 20\% after generative search adoption, with particularly strong reductions in informational and question-based searches \citep{padillaImpactLLMAdoption2025}. This suggests that generative search increasingly acts as a substitute for traditional search engines, particularly for complex or knowledge-seeking queries. 

This shift towards AI search requires marketers to change their performance measurement. Research offers marketers insights into how GEO performance can be measured. \citet{aggarwalGEOGenerativeEngine2024a} indicate that GEO performance can be evaluated using visibility (impression) metrics, such as position-adjusted citation prominence and subjective relevance of sources within generative engine responses. This aligns with \citet{chenGenerativeEngineOptimization2025}, who argue that GEO visibility can be assessed through domain (brand) presence and citation-based source visibility within AI-generated responses.  Therefore, a shift from click-based metrics to visibility occurs \citep{rejon2025generative}. 

However, quantifying visibility as a GEO performance metric presents inherent challenges due to the intransparent architectures of generative search engines. These models function as "black boxes", restricting a firm's ability to predict precisely when or how its brand will be referenced. Whereas SEO visibility typically oscillates along a deterministic ranking spectrum, GEO visibility is subject to far greater instability. LLM-generated responses often exhibit a binary inclusion-exclusion dynamic, where a source is either prominently integrated or omitted entirely. As \citet{wenPositionRisksGenerative2025} articulate, "unlike SEO, which competes for ranked link positions, GEO focuses on inclusion and prominence within LLM-generated answers" \citep[2]{wenPositionRisksGenerative2025}. This phenomenon is driven by the probabilistic nature of token generation and retrieval-augmented evidence selection processes that compress information from diverse sources into a constrained answer space, thereby increasing visibility volatility \citep{aggarwalGEOGenerativeEngine2024a}.

As prior research calls for improved tracking of LLM search outputs and brand visibility within generative engines \citep{wenPositionRisksGenerative2025}, this study extends existing work by focusing not only on the measurement of GEO visibility but also on the stability and consistency of this visibility across prompts and verticals. Thereby, it expands previous GEO visibility papers (e.g., \citep{aggarwalGEOGenerativeEngine2024a}) that do not explicitly measure or empirically quantify GEO visibility fluctuation.

\section{Methods \& Datasets}

For the analysis, two datasets were created. The first dataset contains the daily results of four AI search engines across four Swiss-German campaign verticals, collected over a 45--46-day window (Jan 24 -- Mar 20, 2026). The second contains results of repeated prompts submitted simultaneously on the same day to isolate stochastic variation from temporal drift (see Section~\ref{sec:simul}).

The prompts were derived from high-search-volume SEO keywords. These keywords were entered into Google, and the ``People Also Ask'' feature was subsequently used to identify and generate relevant prompts. Eight prompts per campaign were selected, approximating real user search behaviour, as 70\% of AI-powered search users ask top-of-funnel questions to learn more about products and services \citep{sillimanWinningAgeAI2025}, and users are increasingly moving away from simple keywords toward a conversational tone \citep{martinsEvolutionSEOAge2025}. The study covers four verticals, namely Telecommunications, Real Estate Sales, Sporting Goods, and Consumer Electronics, representing frequently searched domains in Swiss SEO environments.\footnote{Henceforth, these are called campaigns. The original German campaign labels are listed in Appendix \ref{sec:campaign_names}.}For each campaign, eight prompts were entered into four engines: ChatGPT, Gemini, Google AI Mode, and Perplexity. The full list of prompts in both German (original) and English is provided in Appendix~\ref{sec:prompts}. Data coverage is summarised in Table~\ref{tab:data_overview}; rationale for restricting to this single period is given in Section~\ref{sec:limitations}.

\begin{table}[ht]
    \centering
    \small
    \caption{Data coverage by campaign and search engine, Jan 24 -- Mar 20, 2026 (collection days with $\geq 1$ result).}
    \label{tab:data_overview}
    \begin{tabular}{l r r r r r r}
        \toprule
        \textbf{Campaign} & \textbf{Queries} & \textbf{Days} & \textbf{ChatGPT} & \textbf{Gemini} & \textbf{Google AI Mode} & \textbf{Perplexity} \\
        \midrule
        Consumer Electronics        & 8 & 45 & 43 & 22 & 43 & 43 \\
        Real Estate Sales & 8 & 45 & 39 & 26 & 43 & 43 \\
        Sporting Goods      & 8 & 46 & 38 & 23 & 44 & 44 \\
        Telecommunications           & 8 & 45 & 40 & 23 & 43 & 42 \\
        \bottomrule
    \end{tabular}
    \\\small\textit{Note:} Gemini had sporadic gaps in this period. Jan 30, 2026 excluded (citation volume $\approx 2\times$ the daily average).
\end{table}

\subsection*{Similarity Metrics}\label{sec:metrics}

The stability of AI-based search engine results is measured along two dimensions: differences in cited sources and differences in mentioned brands across repeated prompts. Two complementary metrics are used. Jaccard similarity \citep{jaccard1901etude} measures the set overlap of cited sources (or detected brands) between two observations, defined as $J(A,B) = |A \cap B| / |A \cup B|$ (see Appendix~\ref{sec:jaccard_appendix}). It is rank-agnostic but intuitive and easy to interpret. Rank Biased Overlap (RBO, $p{=}0.9$) addresses Jaccard's rank insensitivity by weighting items at the top of the ranked list more heavily than those further down \citep{webberSimilarityMeasureIndefinite2010}, making it appropriate when source position reflects relevance (see Appendix~\ref{sec:rbo_appendix}). The non-extrapolated minimum-bound variant \eqref{eq:rbo-min} is used, truncating the weighted sum at $k = \min(|S|, |T|)$. The unified policy for handling empty source and brand sets is detailed in Appendix \ref{sec:edge_cases}.

\section{Results 1: Source Visibility over Time}
\label{sec:temporal}

\begin{figure}[htb]
    \centering
    \includegraphics[width=1\linewidth]{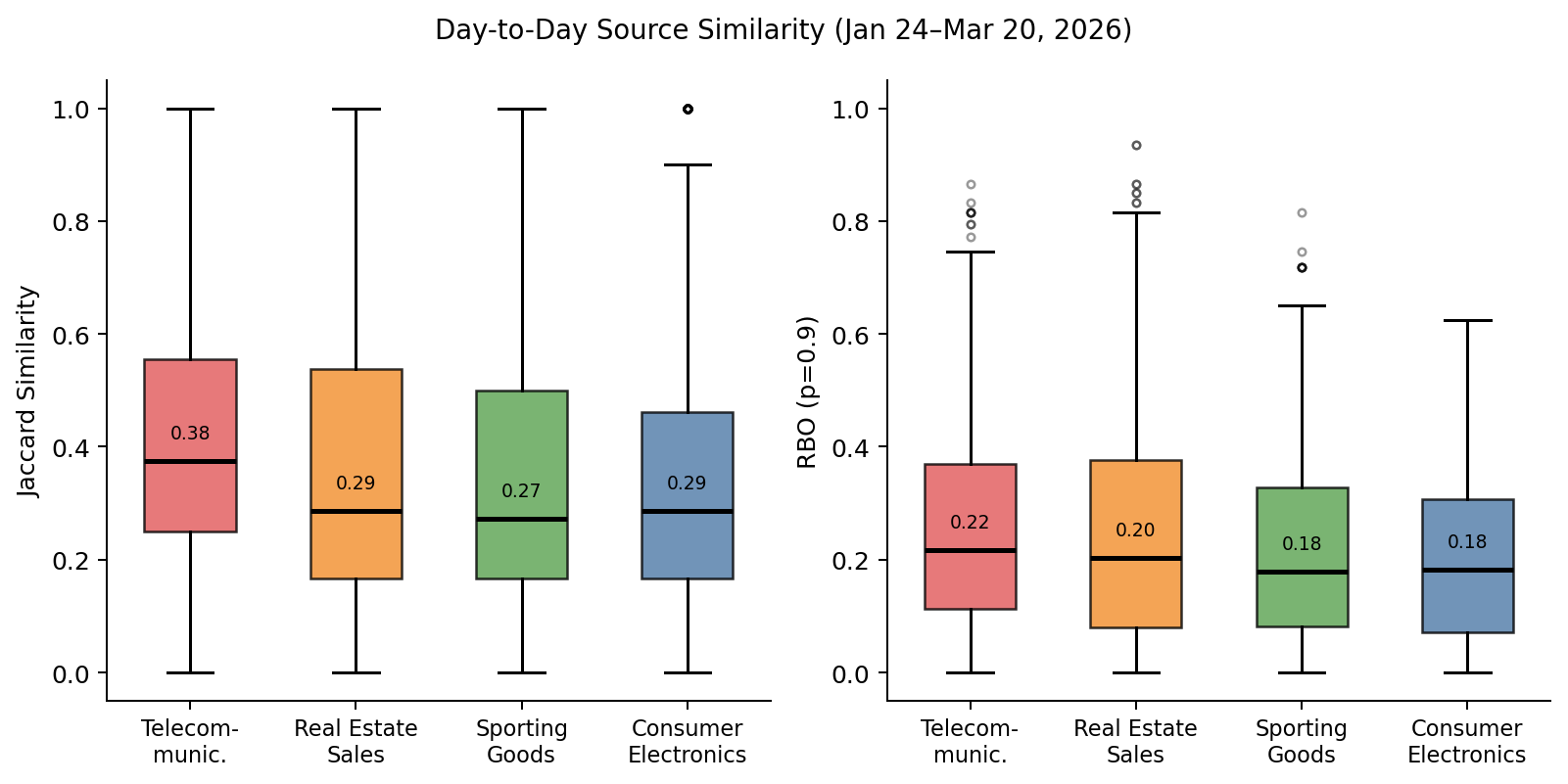}
    \caption{Day-to-day Jaccard similarity (left) and Rank-Biased Overlap (right) for cited sources across all four campaigns (Jan 24 -- Mar 20, 2026). Each box aggregates all 4,044 consecutive-day pairs. Median values annotated above each box. Source sets overlap by only 34--42\% on average.}
    \label{fig:jaccard_rbo}
\end{figure}

Across all four campaigns over the 45--46-day observation window (Jan 24 -- Mar 20, 2026), the day-to-day Jaccard similarity for cited sources averages between 0.34 and 0.42 (see Table~\ref{tab:source_sim_periods} and Figure~\ref{fig:jaccard_rbo}). A Jaccard value of 0.35 implies that, on average, only about 35\% of the cited sources overlap between two consecutive days --- meaning roughly 65\% of all sources change from one day to the next. The RBO scores are consistently lower than Jaccard (0.21--0.26), indicating that not only do the source sets change, but so does the rank order in which they appear. These values confirm that day-to-day source instability is a persistent characteristic of AI search, not a transient artifact.

\begin{table}[ht]
    \centering
    \small
    \caption{Day-to-day source similarity by campaign (Jan 24 -- Mar 20, 2026).}
    \label{tab:source_sim_periods}
    \begin{tabular}{l cccc}
        \toprule
        \textbf{Campaign} & \textbf{Jac.\ Mean} & \textbf{Jac.\ SD} & \textbf{RBO Mean} & \textbf{RBO SD} \\
        \midrule
        Consumer Electronics        & 0.336 & 0.243 & 0.206 & 0.163 \\
        Real Estate Sales & 0.378 & 0.293 & 0.256 & 0.219 \\
        Sporting Goods      & 0.355 & 0.269 & 0.224 & 0.183 \\
        Telecommunications           & 0.423 & 0.244 & 0.253 & 0.180 \\
        \bottomrule
    \end{tabular}
    \\\small\textit{Note:} 4,044 consecutive-day pairs aggregated across all queries and engines. RBO at $p{=}0.9$. Edge-case policy: see Appendix~\ref{sec:edge_cases}.
\end{table}

Beyond source-level instability, we additionally examine brand-level visibility. For each response we detect brands mentioned in the answer text using a campaign-specific lexicon (32--51 canonical brands per vertical; see Appendix~\ref{sec:brand_list} for the full list). Before computing brand similarity, we apply a quality filter: campaigns are included in the brand analysis only if their mean brand-detection rate across all runs exceeds \textbf{70\%}. This threshold was computed on the temporal dataset (Jan 24 -- Mar 20, 2026); the same campaign qualifications were then applied to the simultaneous-run analysis. The Real Estate Sales campaign falls below this threshold (mean detection rate: 53.6\%), driven by several generic tax- and investment-oriented queries (e.g., "Wie viel ist kapitalertragssteuerfrei?") for which LLMs answer without citing any specific brand. It is therefore excluded from brand similarity analyses; its brand lexicon is retained in the appendix for completeness.

Among the three qualifying campaigns (Telecommunications, Sporting Goods, Consumer Electronics), the resulting brand similarity scores, reported in Table~\ref{tab:brand_sim_periods}, are markedly higher than source similarity --- Jaccard values of 0.45--0.59 --- reflecting that brand mentions are somewhat more stable than individual source citations. Nevertheless, substantial day-to-day variation remains: RBO scores of 0.19--0.30 indicate that the implicit ordering of brands within responses also shifts considerably over time. Sporting Goods shows the lowest brand Jaccard (0.45), likely because the large pool of substitutable sports-shoe brands means the model draws from a wide set across days.

\begin{table}[ht]
    \centering
    \small
    \caption{Day-to-day brand similarity by campaign (Jan 24 -- Mar 20, 2026; campaigns with mean brand-detection rate $\geq 70\%$).}
    \label{tab:brand_sim_periods}
    \begin{tabular}{l cccc}
        \toprule
        \textbf{Campaign} & \textbf{Jac.\ Mean} & \textbf{Jac.\ SD} & \textbf{RBO Mean} & \textbf{RBO SD} \\
        \midrule
        Consumer Electronics   & 0.557 & 0.237 & 0.289 & 0.184 \\
        Sporting Goods & 0.453 & 0.326 & 0.187 & 0.182 \\
        Telecommunications      & 0.589 & 0.211 & 0.304 & 0.179 \\
        \bottomrule
    \end{tabular}
    \\\small\textit{Note:} Finance and Real Estate Sales excluded. 2,924 consecutive-day pairs (non-NaN). Edge-case policy: see Appendix~\ref{sec:edge_cases}.
\end{table}

\begin{figure}[htb]
    \centering
    \includegraphics[width=1\linewidth]{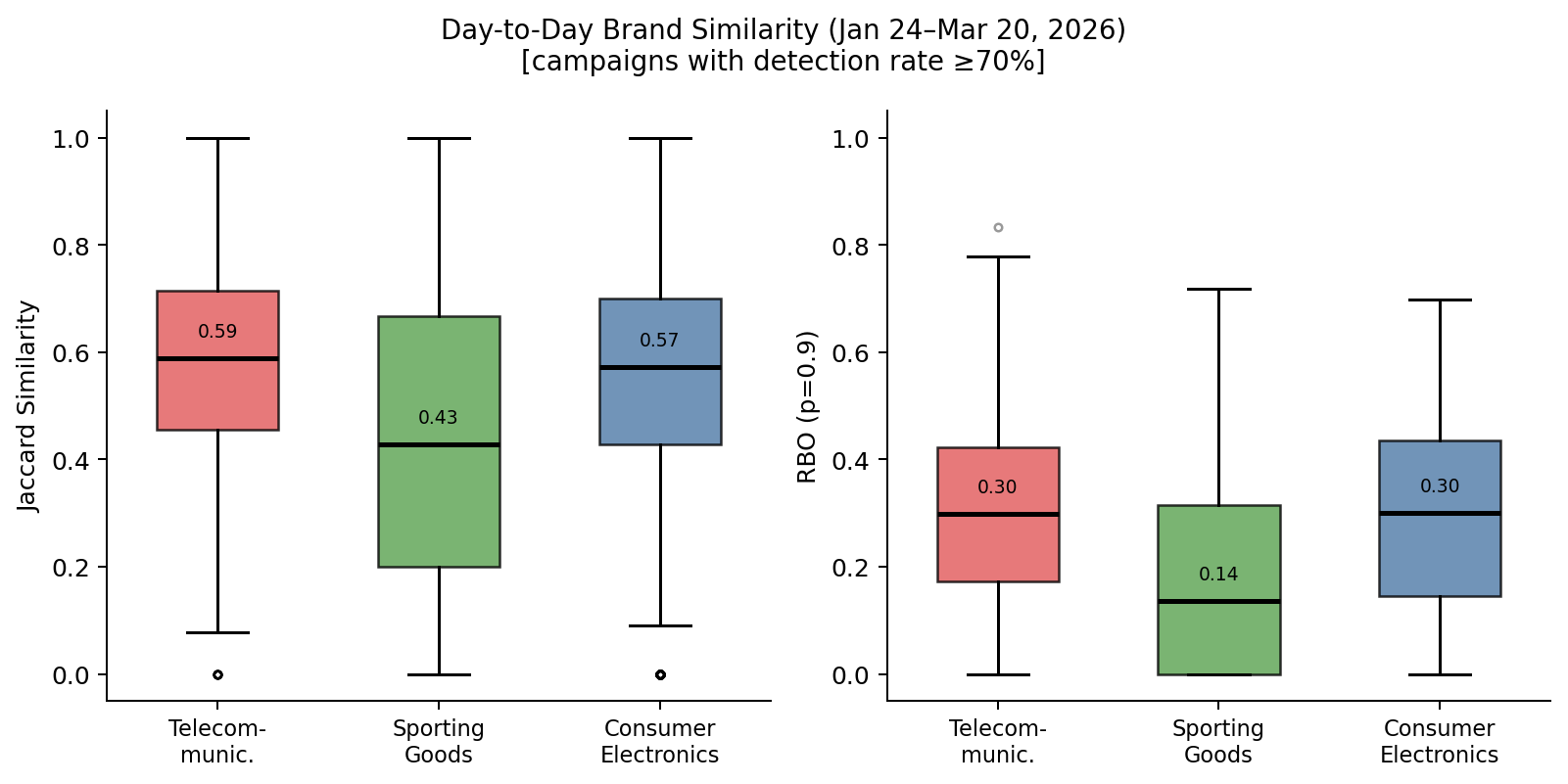}
    \caption{Day-to-day brand similarity (Jaccard, left; RBO, right) for the three campaigns meeting the $\geq 70\%$ brand-detection threshold. Brand stability is higher than source stability but still far from perfect, with broad interquartile ranges indicating substantial response-to-response variation within campaigns. Median values annotated above each box.}
    \label{fig:brand_jaccard_rbo}
\end{figure}

Taken together, these results show that both the specific sources cited and the brands mentioned in AI-generated responses fluctuate substantially from day to day across a 45--46-day window. Source instability appears to be a persistent property of the generative search process rather than an occasional glitch.

\begin{figure}[ht]
    \centering
    \includegraphics[width=\linewidth]{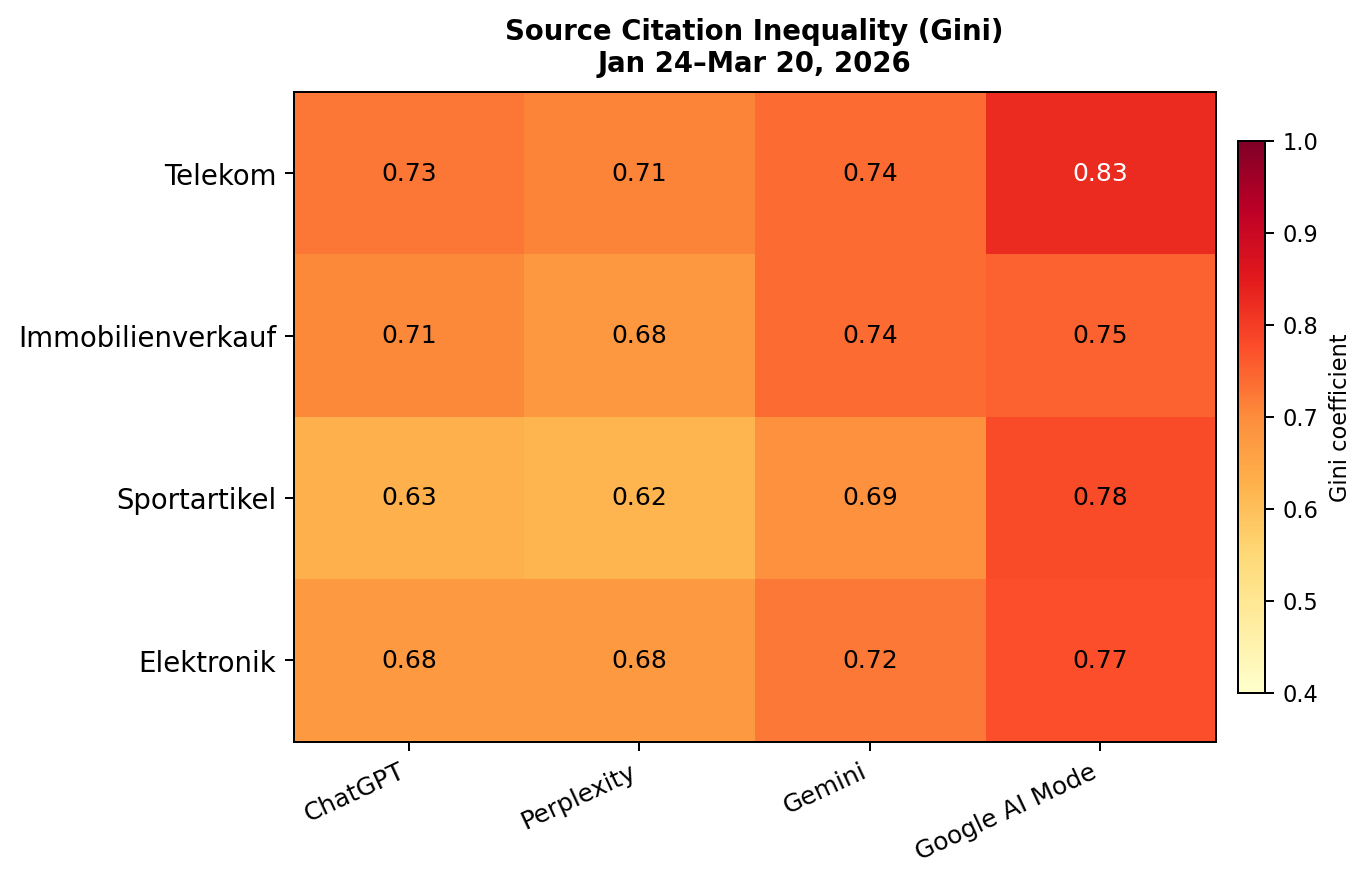}
    \caption{Source citation Gini coefficient by campaign and engine, Jan 24 -- Mar 20, 2026. Values close to 1.0 indicate that a single domain captures nearly all citations; values around 0.6--0.7 indicate moderate concentration. All campaigns and engines exhibit high Gini values (mean 0.715).}
    \label{fig:gini_heatmap}
\end{figure}

Source citation inequality is also notably high: a small number of domains account for the vast majority of citations across all campaigns and engines. Figure~\ref{fig:gini_heatmap} shows Gini coefficients per campaign and engine for Jan 24 -- Mar 20, 2026 (after filtering the \texttt{images.openai.com} CDN artifact; see Section~\ref{sec:limitations}). The mean Gini across all campaigns and engines is 0.715. Google AI Mode exhibits the highest citation concentration (Gini = 0.782), while Perplexity shows the lowest (0.671). Across campaigns, Telecommunications reaches the highest Gini (0.750) and Sporting Goods the lowest (0.680). These values imply a highly unequal citation landscape in which a handful of domains captures most of the AI-generated visibility. The numerical breakdown by campaign and engine and the formula with a worked example are provided in Appendix~\ref{sec:gini_appendix}--\ref{sec:gini_formula}.

What is cited in one run today is not necessarily cited in the same run tomorrow. The drivers of this instability may be external --- algorithmic updates, changes in domain authority, index freshness --- or they may be time-independent, arising from the inherently probabilistic nature of the LLM's output distribution. The simultaneous re-run analysis in Section~\ref{sec:simul} isolates these contributions.

\section{Results 2: Source Visibility Under Simultaneous Re-Runs}
\label{sec:simul}

The temporal analysis in Section~\ref{sec:temporal} establishes that sources and brands change from day to day. However, this variation could in principle be driven by changes external to the model itself --- algorithmic updates or index freshness. To isolate the contribution of the model's inherent stochasticity, we examine cases in which the same prompt was issued multiple times on the \emph{same} calendar day. This removes temporal drift as a confounder: any variation observed within a single day is very likely to originate from the probabilistic nature of the LLM's output or the output system of the respective AI-based search engine.

We draw on a dedicated simultaneous-run collection, which contains 8 prompts per campaign queried up to 10 times to all 4 engines in succession. Because the collection spanned multiple calendar days for some engines (ChatGPT: Mar 21--23, 2026; Perplexity: Mar 23; Gemini: Mar 22--23; Google AI Mode: Mar 24--25), runs are filtered so that only pairs with a timestamp difference of at most 24 hours are compared, ensuring that temporal drift cannot confound the results. For source-overlap analysis a second quality filter is applied: only runs in which the engine returned at least one extracted citation are included, removing zero-citation responses that would otherwise inflate the false-zero Jaccard scores (75.4\% of runs pass this filter; ChatGPT is lowest at 42.2\%, reflecting its tendency to suppress web search on definitional queries). After both filters, 3,409 pairwise source comparisons remain across the four campaigns, with up to 10 runs per engine--prompt group. The similarity edge-case policy described in Section~\ref{sec:metrics} applies here as well. Under classical search-engine assumptions one would expect near-perfect source overlap for identical queries issued within minutes of each other. As Table~\ref{tab:simul_source_jaccard} shows, the actual pairwise Jaccard similarity for sources averages between 0.32 and 0.43 across campaigns --- values in the same range as the day-to-day figures in Section~\ref{sec:temporal}, confirming that intra-day stochastic variation alone accounts for most of the observed instability.

\begin{table}[ht]
    \centering
    \small
    \caption{Pairwise source similarity across repeated runs within a 24-hour window (Jaccard and RBO; runs with at least one extracted citation only).}
    \label{tab:simul_source_jaccard}
    \begin{tabular}{l c c c c c}
        \toprule
        \textbf{Campaign} & \textbf{Pairs} & \textbf{Max Runs} & \textbf{Jaccard Mean} & \textbf{Jaccard SD} & \textbf{RBO Mean} \\
        \midrule
        Consumer Electronics           &   830 &  10 & 0.327 & 0.220 & 0.156 \\
        Real Estate Sales    &   805 &  10 & 0.391 & 0.259 & 0.225 \\
        Sporting Goods         &   886 &  10 & 0.321 & 0.219 & 0.149 \\
        Telecommunications              &   888 &  10 & 0.434 & 0.238 & 0.272 \\
        \bottomrule
    \end{tabular}
    \\\small\textit{Note:} Only pairs with $|\Delta t| \leq 24$ h compared. Runs with zero extracted citations excluded from source analysis. 4 engines: ChatGPT, Perplexity, Gemini, Google AI Mode. Up to 10 reps per engine--prompt. Both-empty source pairs excluded (NaN policy). RBO at $p{=}0.9$.
\end{table}

\begin{figure}[ht]
    \centering
    \includegraphics[width=\linewidth]{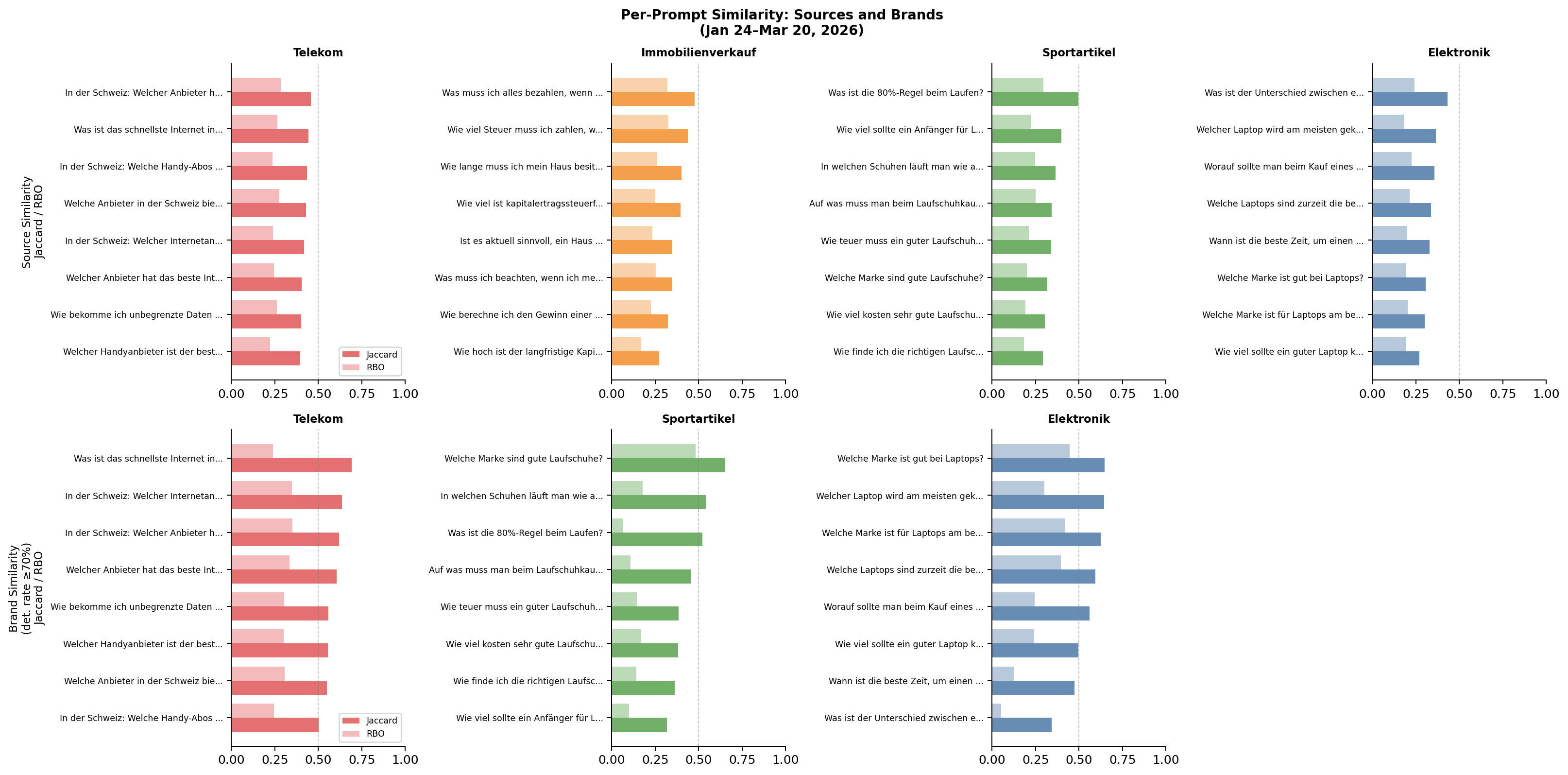}
    \caption{Per-prompt mean Jaccard (dark) and RBO (light) for source similarity (top row, all campaigns) and brand similarity (bottom row, campaigns with detection rate $\geq 70\%$). Prompts are sorted by ascending Jaccard. Vertical dashed line at 0.5 for reference.}
    \label{fig:prompt_analysis}
\end{figure}

We repeat the same pairwise analysis for brand mentions, restricting to the three campaigns that surpass the 70\% detection-rate threshold (Telecommunications, Sporting Goods, Consumer Electronics; see Section~\ref{sec:temporal}). Brand similarity uses all runs with a non-empty response within the 24-hour window (no citation requirement, since brands are extracted from response text). The brand-level Jaccard values (Table~\ref{tab:simul_brand_jaccard}) are higher than source-level values for Consumer Electronics and Telecommunications (0.46--0.48), confirming that brand mentions are somewhat more stable than individual cited sources even within a single day. Sporting Goods shows a lower brand Jaccard (0.33), reflecting the wide interchangeable pool of running-shoe brands from which the model draws. All three campaigns show high within-campaign variance (SD $\approx 0.30$): some prompts yield near-perfect brand consistency across runs while others change almost entirely, consistent with the prompt-level heterogeneity documented in Figure~\ref{fig:prompt_analysis}.

\begin{table}[ht]
    \centering
    \small
    \caption{Pairwise brand similarity across repeated runs within a 24-hour window (Jaccard and RBO; campaigns with mean detection rate $\geq 70\%$).}
    \label{tab:simul_brand_jaccard}
    \begin{tabular}{l c c c c c}
        \toprule
        \textbf{Campaign} & \textbf{Pairs} & \textbf{Max Runs} & \textbf{Jaccard Mean} & \textbf{Jaccard SD} & \textbf{RBO Mean} \\
        \midrule
        Consumer Electronics           &  1235 &  10 & 0.477 & 0.298 & 0.229 \\
        Sporting Goods         &  1027 &  10 & 0.327 & 0.298 & 0.175 \\
        Telecommunications              &  1233 &  10 & 0.463 & 0.301 & 0.220 \\
        \bottomrule
    \end{tabular}
    \\\small\textit{Note:} Finance and Real Estate Sales excluded (detection rate $<70\%$). Pairs with $|\Delta t| \leq 24$ h only (Mar 21--25, 2026; 10 reps). Brand detection uses a campaign-specific lexicon (see Appendix~\ref{sec:brand_list}). Both-empty pairs excluded per edge-case policy (Section~\ref{sec:metrics}).
\end{table}

\begin{table}[ht]
    \centering
    \small
    \caption{Mean pairwise similarity within 24 hours by engine: sources (cited runs only, all 4 campaigns) and brands (Telecommunications, Sporting Goods, Consumer Electronics).}
    \label{tab:simul_by_engine}
    \begin{tabular}{l cccc}
        \toprule
        & \multicolumn{2}{c}{\textbf{Source}} & \multicolumn{2}{c}{\textbf{Brand}} \\
        \cmidrule(lr){2-3}\cmidrule(lr){4-5}
        \textbf{Engine} & \textbf{Jac.\ Mean} & \textbf{RBO Mean} & \textbf{Jac.\ Mean} & \textbf{RBO Mean} \\
        \midrule
        ChatGPT              & 0.233 & 0.088 & 0.437 & 0.192 \\
        Perplexity           & 0.282 & 0.102 & 0.492 & 0.202 \\
        Gemini               & 0.505 & 0.230 & 0.409 & 0.196 \\
        Google AI Mode       & 0.318 & 0.254 & 0.375 & 0.238 \\
        \bottomrule
    \end{tabular}
    \\\small\textit{Note:} Source columns include only runs with $\geq 1$ extracted citation and pairs with $|\Delta t| \leq 24$ h. Brand columns restricted to campaigns meeting the $\geq 70\%$ detection-rate threshold. RBO at $p{=}0.9$.
\end{table}

\begin{figure}[ht]
    \centering
    \includegraphics[width=\linewidth]{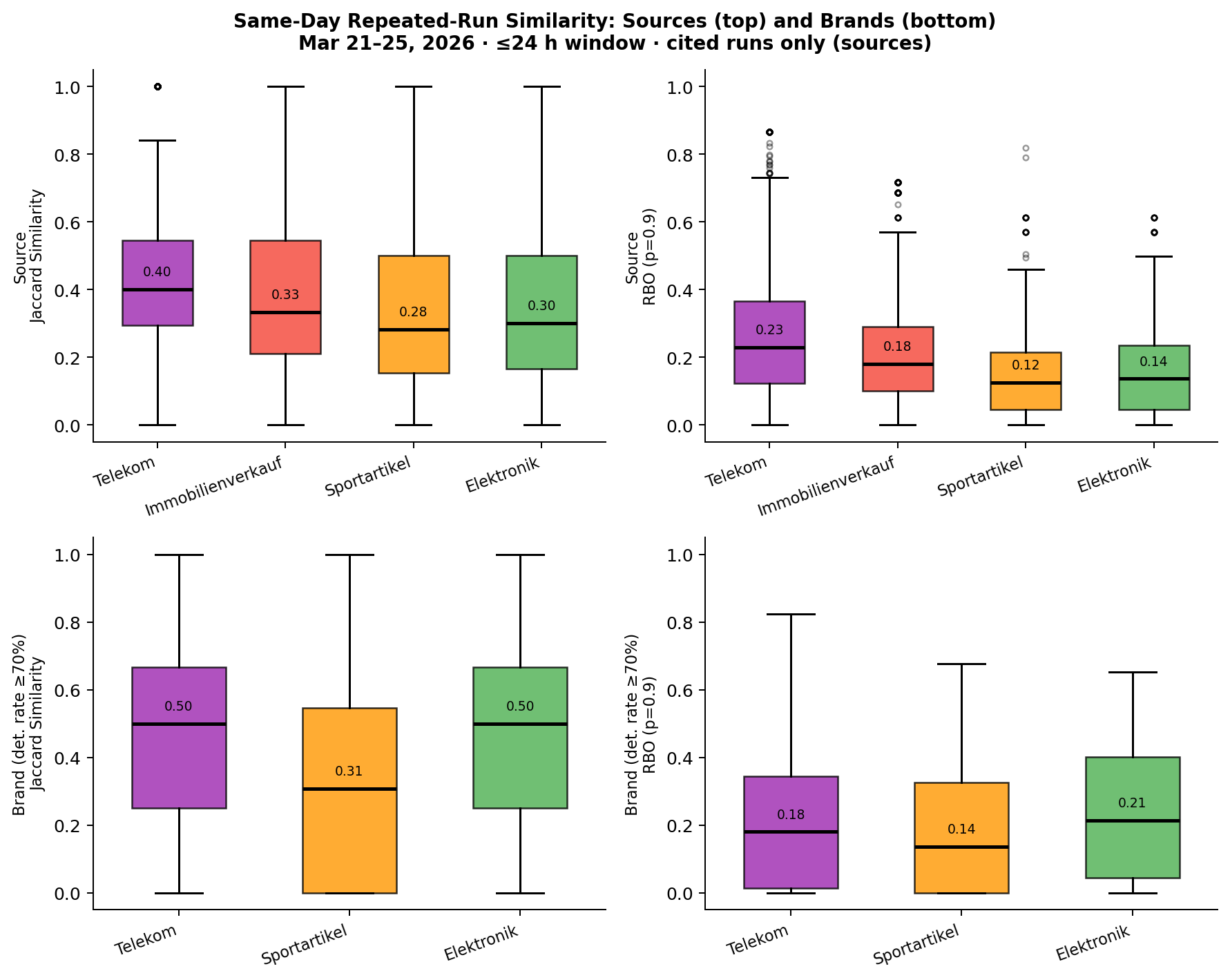}
    \caption{Pairwise Jaccard and RBO for sources (top row) and brands (bottom row) across repeated runs within a 24-hour window (geo-brand-monitor collection, Mar 21--25, 2026; 4 engines, up to 10 reps). Source boxes include only runs with $\geq 1$ extracted citation. Source overlap averages 32--43\%; brand overlap ranges from 33\% (Sporting Goods) to 48\% (Consumer Electronics). Median values annotated above each box.}
    \label{fig:simul_boxplot}
\end{figure}

Figure~\ref{fig:simul_boxplot} visualises the full distribution of pairwise Jaccard scores across campaigns. The consistently low median values and broad interquartile ranges demonstrate that LLM output variation is not reducible to external temporal factors: a substantial fraction of observed instability originates from the model's stochastic generation process itself.

The practical implication is direct: if a marketer queries an AI search engine once on a given day, the resulting brand-visibility snapshot may differ substantially from a second query executed minutes later under identical conditions. Characterising true GEO visibility therefore requires aggregating over multiple runs rather than relying on a single observation.

Figure~\ref{fig:prompt_analysis} shows per-prompt mean Jaccard and RBO values for both source and brand similarity across all campaigns. The top row covers source similarity for all four campaigns; the bottom row covers brand similarity for the three qualifying campaigns (Telecommunications, Sporting Goods, Consumer Electronics). Within each campaign, prompts vary substantially in their similarity levels, suggesting that query specificity --- rather than engine behaviour alone --- determines how consistently a prompt is answered. Specific product queries (e.g., ``Welche Sportschuhe sind die besten?'') tend to attract more consistent source and brand sets than broad, generic queries.

As Figure~\ref{fig:prompt_analysis} illustrates, variability across prompts is substantial: some prompts yield consistently high similarity (Jaccard $>0.8$) while others remain persistently low ($<0.2$). This prompt-level heterogeneity implies that single-prompt-based visibility assessments are unreliable, and that monitoring strategies must account for query-level variation in addition to temporal and simultaneous-run variation. Together with the temporal results, this strongly motivates a repeated-measurement framework for GEO monitoring.

\section{Limitations}
\label{sec:limitations}

Several data-quality and methodological limitations should be noted when interpreting the results.

\textbf{ChatGPT data} collected contained \texttt{images.openai.com} --- an OpenAI image-delivery CDN --- as a spurious domain data (889 occurrences, 5.8\% of ChatGPT P2 citations). Because mixing API and interface data would create a methodological inconsistency for ChatGPT specifically, all analyses in this paper are restricted to the January 24 -- March 20, 2026 window, where collection is consistent across all four engines. The \texttt{images.openai.com} domain is additionally filtered from all calculations. Future studies should pin collection method and model version across the full observation window.

\textbf{Swiss-server context.} All data were collected from servers located in Switzerland. Prompts are therefore served with Swiss IP addresses and locale settings, which may affect geo-personalised index selection, language weighting, and citation patterns. Results may not generalise to other regional or linguistic markets.

\textbf{Simultaneous-run collection window.} The geo-brand-monitor simultaneous collection spanned multiple calendar days for some engines (ChatGPT: Mar 21--23; Perplexity: Mar 23; Gemini: Mar 22--23; Google AI Mode: Mar 24--25, 2026). The raw dataset additionally contains responses from Google AI Overviews (``Google AIO'', Mar 22--23), a distinct Google product that generates AI-summary snippets within regular search results rather than operating as a dedicated AI search interface. Because Google AIO differs fundamentally in interaction mode and citation behaviour from the four dedicated AI search engines in the study, it is excluded from all analyses; the study focuses on ChatGPT, Gemini, Google AI Mode, and Perplexity. The analysis further controls for multi-day collection by including only pairs whose timestamps are within 24 hours of each other; all cross-day pairs exceeding this window are discarded. Additionally, ChatGPT activates web search only for specific queries, leaving 57.8\% of its runs with zero citations; the source-similarity analysis therefore excludes zero-citation runs across all engines to ensure comparisons reflect genuine source overlap rather than collection failures.

\textbf{Brand-detection coverage.} Brand detection relies on substring matching of a fixed lexicon. Brands cited via synonyms, abbreviations, or paraphrases are missed; conversely, generic terms that are substrings of brand names may produce false positives. The 70\% detection-rate threshold used to qualify campaigns for brand similarity analysis mitigates the worst of this, but does not eliminate the problem.

\section{Conclusion}
The research demonstrates that visibility in AI search is inherently unstable and cannot be treated like traditional SEO rankings. In contrast to SEO, where results may shift in position but typically remain present in the ranking set, generative search operates on an inclusion--exclusion dynamic in which brands or sources may appear in one response and disappear entirely in another. Even when identical prompts are executed simultaneously under controlled conditions, cited sources and brand mentions vary substantially. Source sets overlap by only 34--42\% between consecutive days; brand sets somewhat more, at 45--59\%, but with wide variance. As a result, single observations of AI visibility are misleading and risk over- or underestimating true brand presence.

Instead of relying on snapshot metrics, marketers must conceptualize GEO performance as the probability of being mentioned across repeated runs. This shift from deterministic rankings to probabilistic visibility fundamentally changes how marketing performance in AI search should be monitored and managed. Several concrete implications follow:

\begin{itemize}
    \item \textbf{Minimum run count.} A single daily query cannot provide a reliable estimate of true visibility. A bootstrap convergence analysis on the 10-run simultaneous dataset (Appendix~\ref{sec:convergence}) shows that the standard error of the estimated per-brand detection rate drops below 0.10 at $n = 7$ runs (95\% CI $\pm 0.158$) and below 0.08 at $n = 8$ runs (95\% CI $\pm 0.121$). For source coverage the convergence is slower: SE $< 0.10$ requires $n = 8$ runs, reflecting higher source-level stochasticity. Practitioners should therefore use at least 7 runs per prompt per day for brand visibility monitoring, and at least 8 runs when source-level coverage matters.
    \item \textbf{Multi-prompt coverage.} Prompt-level Jaccard scores vary widely within campaigns (from below 0.2 to above 0.8). Monitoring based on one or two prompts will reflect the idiosyncrasies of those prompts rather than campaign-level visibility. A large prompt portfolio of diverse queries is advisable.
    
    \item \textbf{Sustained observation windows.} Day-to-day source instability ($\approx 65\%$ turnover) means short observation windows, e.g. days to a week, are insufficient to distinguish signal from noise. A per-brand rolling-window convergence analysis on the temporal dataset (Appendix~\ref{sec:temporal_convergence}) shows that the standard error of a $d$-day per-brand detection rate estimate drops below 0.10 at $d = 10$ days and below 0.05 at $d = 24$ days (95\% CI $\pm 0.105$ at $d = 21$; $\pm 0.065$ at $d = 28$). Short windows that appear precise at the campaign level are substantially noisier when tracking individual brands. Rolling aggregation over two to four weeks is therefore recommended to obtain per-brand estimates that are both statistically stable and representative of sustained visibility rather than momentary snapshots.
    \item \textbf{Campaign-specific benchmarking.} Citation concentration (Gini $\approx 0.71$ on average) varies meaningfully across both campaigns and engines. Google AI Mode concentrates citations most strongly; Perplexity distributes them most evenly. Marketers should set engine-specific visibility baselines rather than applying a single threshold across all AI search products.
    \item \textbf{Brand vs.\ source monitoring.} Brand-level day-to-day stability (Jaccard 0.45--0.59) exceeds source-level stability (0.34--0.42), suggesting that brand presence aggregated over a campaign is a more reliable KPI than tracking individual cited URLs. Source-level monitoring remains valuable for understanding which content pieces drive inclusion.
\end{itemize}

Future work should examine whether the instability patterns documented here hold across other languages and regional markets, and whether targeted GEO interventions (e.g., structured content, authoritative backlink profiles) can shift a brand's inclusion probability in a measurable and durable way.

\bibliographystyle{abbrvnat}
\clearpage
\bibliography{references_processed}  %%% Uncomment this line and comment out the ``thebibliography'' section below to use the external .bib file (using bibtex) .

@online{aggarwalGEOGenerativeEngine2024a,
  title = {{{GEO}}: {{Generative Engine Optimization}}},
  shorttitle = {{{GEO}}},
  author = {Aggarwal, Pranjal and Murahari, Vishvak and Rajpurohit, Tanmay and Kalyan, Ashwin and Narasimhan, Karthik and Deshpande, Ameet},
  date = {2024-06-28},
  year = {2024},
  eprint = {2311.09735},
  eprinttype = {arXiv},
  eprintclass = {cs},
  doi = {10.48550/arXiv.2311.09735},
  url = {http://arxiv.org/abs/2311.09735},
  urldate = {2026-02-06},
  langid = {english},
  pubstate = {prepublished}
}

@online{AIVisibilitySEO,
  year = {2025},
  title = {AI Visibility: SEO, AEO \& GEO für digitale Sichtbarkeit | Deloitte Deutschland},
  shorttitle = {AI Visibility},
  author = {Krisinger, Jens},
  url = {https://www.deloitte.com/de/de/services/consulting/perspectives/ai-visibility.html},
  urldate = {2026-02-21},
  langid = {german}
}

@report{chatterjiHowPeopleUse2025a,
  title = {How People Use Chatgpt},
  author = {Chatterji, Aaron and Cunningham, Thomas and Deming, David J and Hitzig, Zoe and Ong, Christopher and Shan, Carl Yan and Wadman, Kevin},
  date = {2025},
  year = {2025},
  institution = {National Bureau of Economic Research}
}

@online{chenEvaluatingLargeLanguage2021,
  year = {2021},
  title = {Evaluating {{Large Language Models Trained}} on {{Code}}},
  author = {Chen, Mark and Tworek, Jerry and Jun, Heewoo and Yuan, Qiming and Pinto, Henrique Ponde de Oliveira and Kaplan, Jared and Edwards, Harri and Burda, Yuri and Joseph, Nicholas and Brockman, Greg and Ray, Alex and Puri, Raul and Krueger, Gretchen and Petrov, Michael and Khlaaf, Heidy and Sastry, Girish and Mishkin, Pamela and Chan, Brooke and Gray, Scott and Ryder, Nick and Pavlov, Mikhail and Power, Alethea and Kaiser, Lukasz and Bavarian, Mohammad and Winter, Clemens and Tillet, Philippe and Such, Felipe Petroski and Cummings, Dave and Plappert, Matthias and Chantzis, Fotios and Barnes, Elizabeth and Herbert-Voss, Ariel and Guss, William Hebgen and Nichol, Alex and Paino, Alex and Tezak, Nikolas and Tang, Jie and Babuschkin, Igor and Balaji, Suchir and Jain, Shantanu and Saunders, William and Hesse, Christopher and Carr, Andrew N. and Leike, Jan and Achiam, Josh and Misra, Vedant and Morikawa, Evan and Radford, Alec and Knight, Matthew and Brundage, Miles and Murati, Mira and Mayer, Katie and Welinder, Peter and McGrew, Bob and Amodei, Dario and McCandlish, Sam and Sutskever, Ilya and Zaremba, Wojciech},
  date = {2021},
  year = {2021},
  doi = {10.48550/ARXIV.2107.03374},
  url = {https://arxiv.org/abs/2107.03374},
  urldate = {2026-04-04},
  pubstate = {prepublished},
  version = {2}
}

@online{chenGenerativeEngineOptimization2025,
  title = {Generative {{Engine Optimization}}: {{How}} to {{Dominate AI Search}}},
  shorttitle = {Generative {{Engine Optimization}}},
  author = {Chen, Mahe and Wang, Xiaoxuan and Chen, Kaiwen and Koudas, Nick},
  date = {2025-09-10},
  year = {2025},
  eprint = {2509.08919},
  eprinttype = {arXiv},
  eprintclass = {cs},
  doi = {10.48550/arXiv.2509.08919},
  url = {http://arxiv.org/abs/2509.08919},
  urldate = {2026-02-06},
  langid = {english},
  pubstate = {prepublished}
}

@article{economistHowAIDisrupting,
  year = {2025},
  entrysubtype = {magazine},
  title = {How {{AI}} Is Disrupting Shopping},
  author = {Economist, The},
  journaltitle = {The Economist},
  issn = {0013-0613},
  url = {https://www.economist.com/business/2025/12/08/how-ai-is-disrupting-shopping},
  urldate = {2026-02-06}
}

@book{efronIntroductionBootstrap1994,
  title = {An {{Introduction}} to the {{Bootstrap}}},
  author = {Efron, Bradley and Tibshirani, R.J.},
  date = {1994-05-15},
  year = {1994},
  edition = {0},
  publisher = {{Chapman and Hall/CRC}},
  doi = {10.1201/9780429246593},
  url = {https://www.taylorfrancis.com/books/9781000064988},
  urldate = {2026-04-04},
  isbn = {978-0-429-24659-3},
  langid = {english}
}

@article{giniMeasurementInequalityIncomes1921,
  title = {Measurement of {{Inequality}} of {{Incomes}}},
  author = {Gini, Corrado},
  date = {1921-03},
  year = {1921},
  journaltitle = {The Economic Journal},
  shortjournal = {The Economic Journal},
  volume = {31},
  number = {121},
  eprint = {10.2307/2223319},
  eprinttype = {jstor},
  pages = {124},
  issn = {00130133},
  doi = {10.2307/2223319},
  url = {https://www.jstor.org/stable/10.2307/2223319?origin=crossref},
  urldate = {2026-04-04}
}

@online{goodwinGooglesSearchMarket2025,
  title = {Google's Search Market Share Drops below 90\% for First Time since 2015},
  author = {Goodwin, Danny},
  date = {2025-01-13T16:21:45+00:00},
  year = {2025},
  url = {https://searchengineland.com/google-search-market-share-drops-2024-450497},
  urldate = {2026-02-06},
  langid = {english},
  organization = {Search Engine Land}
}

@article{jaccard1901etude,
  title = {Étude Comparative de La Distribution Florale Dans Une Portion Des {{Alpes}} et Des {{Jura}}},
  author = {Jaccard, Paul},
  date = {1901},
  year = {1901},
  journaltitle = {Bull Soc Vaudoise Sci Nat},
  volume = {37},
  pages = {547--579}
}

@incollection{Leskovec_Rajaraman_Ullman_2014,
  title = {Finding Similar Items},
  booktitle = {Mining of Massive Datasets},
  author = {Leskovec, Jure and Rajaraman, Anand and Ullman, Jeffrey David},
  date = {2014},
  year = {2014},
  pages = {68--122},
  publisher = {Cambridge University Press},
  location = {Cambridge}
}

@online{liorReliableEvalRecipeStochastic2025,
  title = {{{ReliableEval}}: {{A Recipe}} for {{Stochastic LLM Evaluation}} via {{Method}} of {{Moments}}},
  shorttitle = {{{ReliableEval}}},
  author = {Lior, Gili and Habba, Eliya and Levy, Shahar and Caciularu, Avi and Stanovsky, Gabriel},
  date = {2025},
  year = {2025},
  doi = {10.48550/ARXIV.2505.22169},
  url = {https://arxiv.org/abs/2505.22169},
  urldate = {2026-04-04},
  pubstate = {prepublished},
  version = {2}
}

@article{martinsEvolutionSEOAge2025,
  title = {The {{Evolution}} of {{SEO}} in the {{Age}} of {{Generative Search Engines}}},
  author = {Martins, João Maria Gibert Prates de Oliveira},
  date = {2025-10-28},
  year = {2025},
  eprint = {10362/190409},
  eprinttype = {hdl},
  url = {http://hdl.handle.net/10362/190409},
  urldate = {2026-02-06},
  langid = {english}
}

@article{mizrahiStateWhatArt2024,
  title = {State of {{What Art}}? {{A Call}} for {{Multi-Prompt LLM Evaluation}}},
  shorttitle = {State of {{What Art}}?},
  author = {Mizrahi, Moran and Kaplan, Guy and Malkin, Dan and Dror, Rotem and Shahaf, Dafna and Stanovsky, Gabriel},
  date = {2024-08-01},
  year = {2024},
  journaltitle = {Transactions of the Association for Computational Linguistics},
  volume = {12},
  pages = {933--949},
  issn = {2307-387X},
  doi = {10.1162/tacl_a_00681},
  url = {https://direct.mit.edu/tacl/article/doi/10.1162/tacl_a_00681/123885/State-of-What-Art-A-Call-for-Multi-Prompt-LLM},
  urldate = {2026-04-04},
  langid = {english}
}

@online{padillaImpactLLMAdoption2025,
  type = {SSRN Scholarly Paper},
  title = {The {{Impact}} of {{LLM Adoption}} on {{Online User Behavior}}},
  author = {Padilla, Nicolas and Lam, H. Tai and Lambrecht, Anja and Hollenbeck, Brett},
  date = {2025-12-24},
  year = {2025},
  number = {5393256},
  eprint = {5393256},
  eprinttype = {Social Science Research Network},
  location = {Rochester, NY},
  doi = {10.2139/ssrn.5393256},
  url = {https://papers.ssrn.com/abstract=5393256},
  urldate = {2026-02-06},
  langid = {english},
  pubstate = {prepublished}
}

@incollection{rejon2025generative,
  title = {Generative Engine Optimization: {{How}} Search Engines Integrate {{AI-generated}} Content into Conventional Queries},
  booktitle = {Encyclopedia of Artificial Intelligence in Marketing},
  author = {Rejón-Guardia, Francisco and Molinillo, Sebastián and Anaya-Sánchez, Rafael},
  date = {2025},
  year = {2025},
  pages = {1--8},
  publisher = {Springer}
}

@online{sillimanWinningAgeAI2025,
  title = {Winning in the Age of {{AI}} Search | {{McKinsey}}},
  author = {Silliman, Elizabeth and Boudet, Julien and Robinson, Kelsey},
  date = {2025-10-16},
  year = {2025},
  url = {https://www.mckinsey.com/capabilities/growth-marketing-and-sales/our-insights/new-front-door-to-the-internet-winning-in-the-age-of-ai-search?utm_source=chatgpt.com},
  urldate = {2026-02-06}
}

@article{sturzeAgileMarketingPerformance2022,
  title = {Agile {{Marketing Performance Management}}},
  author = {Stürze, Sascha and Hoyer, Markus and Righetti, Claudio and Rasztar, Matthias},
  date = {2022},
  year = {2022},
  journaltitle = {Management for Professionals},
  publisher = {Springer}
}

@article{webberSimilarityMeasureIndefinite2010,
  title = {A Similarity Measure for Indefinite Rankings},
  author = {Webber, William and Moffat, Alistair and Zobel, Justin},
  date = {2010-11},
  year = {2010},
  journaltitle = {ACM Transactions on Information Systems},
  shortjournal = {ACM Trans. Inf. Syst.},
  volume = {28},
  number = {4},
  pages = {1--38},
  issn = {1046-8188, 1558-2868},
  doi = {10.1145/1852102.1852106},
  url = {https://dl.acm.org/doi/10.1145/1852102.1852106},
  urldate = {2025-11-26},
  langid = {english}
}

@online{wenPositionRisksGenerative2025,
  title = {Position: {{On}} the {{Risks}} of {{Generative Engine Optimization}} in the {{Era}} of {{LLMs}}},
  shorttitle = {Position},
  author = {Wen, Yizhu and Zhang, Nan and Yuan, Haohan and Chen, Xun and Zhang, Haopeng and Guo, Hanqing},
  date = {2025-12-20},
  year = {2025},
  eprinttype = {Preprints},
  doi = {10.36227/techrxiv.176620816.64043115/v1},
  url = {https://www.techrxiv.org/users/1010717/articles/1370817-position-on-the-risks-of-generative-engine-optimization-in-the-era-of-llms?commit=d9d611ded6e17d7b0dc93e855718f984775127dd},
  urldate = {2026-02-06},
  langid = {english},
  pubstate = {prepublished}
}

%%% Uncomment this section and comment out the \bibliography{references} line above to use inline references.
% \begin{thebibliography}{1}

% 	\bibitem{kour2014real}
% 	George Kour and Raid Saabne.
% 	\newblock Real-time segmentation of on-line handwritten arabic script.
% 	\newblock In {\em Frontiers in Handwriting Recognition (ICFHR), 2014 14th
% 			International Conference on}, pages 417--422. IEEE, 2014.

% 	\bibitem{kour2014fast}
% 	George Kour and Raid Saabne.
% 	\newblock Fast classification of handwritten on-line arabic characters.
% 	\newblock In {\em Soft Computing and Pattern Recognition (SoCPaR), 2014 6th
% 			International Conference of}, pages 312--318. IEEE, 2014.

% 	\bibitem{hadash2018estimate}
% 	Guy Hadash, Einat Kermany, Boaz Carmeli, Ofer Lavi, George Kour, and Alon
% 	Jacovi.
% 	\newblock Estimate and replace: A novel approach to integrating deep neural
% 	networks with existing applications.
% 	\newblock {\em arXiv preprint arXiv:1804.09028}, 2018.

% \end{thebibliography}

\newpage
\appendix

\section{Campaign Names}
\label{sec:campaign_names}

The four study verticals are reflecting German terminology. Table~\ref{tab:campaign_names} maps the English names used throughout this paper to the original German labels.

\begin{table}[ht]
    \centering
    \small
    \caption{English campaign names used in this paper and corresponding original German labels.}
    \label{tab:campaign_names}
    \begin{tabular}{ll}
        \toprule
        \textbf{English (used in paper)} & \textbf{Original German label} \\
        \midrule
        Telecommunications   & Telekom \\
        Real Estate Sales    & Immobilienverkauf \\
        Sporting Goods       & Sportartikel \\
        Consumer Electronics & Elektronik \\
        \bottomrule
    \end{tabular}
\end{table}

\section{Dataset --- Visibility over Time}

\begin{table}[ht]
    \centering
    \small
    \caption{Full data coverage by campaign and engine, Jan 24 -- Mar 20, 2026 (collection days with $\geq 1$ result). Finance excluded; all data collected via web-interface scraping.}
    \label{tab:full_data_availability}
    \begin{tabular}{l r r r r r r}
        \toprule
        \textbf{Campaign} & \textbf{Queries} & \textbf{Days} & \textbf{ChatGPT} & \textbf{Gemini} & \textbf{Google AI Mode} & \textbf{Perplexity} \\
        \midrule
        Consumer Electronics        & 8 & 45 & 43 & 22 & 43 & 43 \\
        Real Estate Sales & 8 & 45 & 39 & 26 & 43 & 43 \\
        Sporting Goods      & 8 & 46 & 38 & 23 & 44 & 44 \\
        Telecommunications           & 8 & 45 & 40 & 23 & 43 & 42 \\
        \bottomrule
    \end{tabular}
    \\\small\textit{Note:} Gemini had sporadic gaps. Jan 30, 2026 excluded (citation volume $\approx 2\times$ daily average). \texttt{images.openai.com} filtered from all calculations.
\end{table}

\section{Similarity Edge-Case Policy}
\label{sec:edge_cases}

The following policy is applied uniformly to both source and brand similarity calculations throughout the paper.

\begin{itemize}
    \item \textbf{Both lists empty}: the pair is excluded from aggregation (assigned NaN). When both runs return the same empty state --- no sources cited or no brands detected --- there is agreement, but it carries no information about \emph{which} items are stably cited. Including such pairs would inflate mean similarity scores, particularly for campaigns or queries with low detection rates.
    \item \textbf{One list empty, the other non-empty}: Jaccard $= 0.0$, RBO $= 0.0$ (maximum disagreement). One run produced citations or brand mentions; the other did not. This is treated as the most severe form of instability.
    \item \textbf{Both lists non-empty}: standard Jaccard and RBO as defined in Appendix~\ref{sec:jaccard_appendix} and \ref{sec:rbo_appendix}.
\end{itemize}

This policy is especially important for brand similarity: campaigns with low detection rates (notably Real Estate Sales, mean 53.6\%) would otherwise exhibit inflated Jaccard values driven by many (empty, empty) run-pairs. The 70\% detection-rate threshold for including a campaign in brand similarity analyses further mitigates this issue.

\section{Jaccard Similarity}
\label{sec:jaccard_appendix}

The Jaccard Similarity (also known as the Jaccard Index) measures the similarity between two finite sample sets \citep{jaccard1901etude}. It is defined as the size of the intersection divided by the size of the union of the sample sets. It is widely used in information retrieval and biology to compare the overlap of two unweighted sets.

Given two sets $A$ and $B$, the Jaccard coefficient $J(A, B)$ is defined as \citep{jaccard1901etude,Leskovec_Rajaraman_Ullman_2014}:

\begin{equation}
    J(A, B) = \frac{|A \cap B|}{|A \cup B|} = \frac{|A \cap B|}{|A| + |B| - |A \cap B|}
\end{equation}

\noindent Where:
\begin{itemize}
    \item $0 \le J(A, B) \le 1$.
    \item If the sets are disjoint ($A \cap B = \emptyset$), $J = 0$.
    \item If the sets are identical ($A = B$), $J = 1$.
\end{itemize}

\section{Rank Biased Overlap (RBO)}
\label{sec:rbo_appendix}

Rank Biased Overlap is a similarity measure for indefinite rankings. Unlike the Jaccard index, RBO is designed for \textit{ranked lists} rather than sets. It weights items at the top of the list more heavily than those at the bottom and can handle lists of different lengths or lists that are not conjoint (do not contain the same items).

RBO calculates similarity based on the overlap at each depth $d$, weighted by a geometric decay determined by a persistence parameter $p$. The general definition sums to infinite depth \citep{webberSimilarityMeasureIndefinite2010}:

\begin{equation}
    \mathrm{RBO}(S, T, p) = (1 - p) \sum_{d=1}^{\infty} p^{d-1} \cdot A_d
\end{equation}

\noindent Where:
\begin{itemize}
    \item $S$ and $T$ are the two ranked lists.
    \item $p$ is the persistence parameter ($0 < p < 1$). A higher $p$ indicates a stronger interest in the lower-ranked items (the ``tail'' of the list).
    \item $d$ is the rank depth.
    \item $A_d$ is the agreement (overlap) at depth $d$, calculated as:
    \begin{equation}
        A_d = \frac{|S_{:d} \cap T_{:d}|}{d}
    \end{equation}
    Here, $S_{:d}$ and $T_{:d}$ denote the sets of items present in lists $S$ and $T$ up to rank $d$.
\end{itemize}

\textbf{Implementation note.} Because both source and brand lists are finite, the infinite sum must be truncated in practice. This study uses the \emph{non-extrapolated} (minimum-bound) variant, in which the sum is truncated at $k = \min(|S|, |T|)$ --- the length of the shorter list --- and $A_d = 0$ is assumed for all $d > k$. This gives:

\begin{equation}
    \mathrm{RBO}_{\min}(S, T, p) = (1 - p) \sum_{d=1}^{k} p^{d-1} \cdot A_d, \quad k = \min(|S|, |T|)
    \label{eq:rbo-min}
\end{equation}

For identical lists of length $k$, this yields $1 - p^{k}$ rather than 1.0, because the geometric series is not summed to infinity. At $p = 0.9$ and $k = 5$, for instance, $\mathrm{RBO}_{\min} = 1 - 0.9^5 \approx 0.41$ for perfectly matching lists. This is a \emph{conservative lower bound} on the true RBO: any unobserved overlap beyond depth $k$ would only increase the score. The minimum-bound variant is appropriate here because the lists under comparison (AI-generated source citations or brand detections) vary in length across runs, and extrapolating beyond what was actually observed introduces assumptions that are not warranted. Duplicate items are removed from each list before computation to satisfy the package's requirement for unique elements.

While Jaccard is set-based and order-agnostic, RBO is rank-sensitive. Jaccard is appropriate when the presence of an item is the only factor, whereas RBO is appropriate when the position of the item signifies importance (e.g., search engine results). RBO scores are computed using the Python implementation by Changyao Chen.\footnote{The respective Python package can be found \href{https://github.com/changyaochen/rbo}{here}.}

\section{Brand Lexicon}
\label{sec:brand_list}

The brand lexicon can be found in Table \ref{tab:brand_lexicon}.
\begin{table}[H]
    \centering
    \small
    \caption{Complete brand lexicon used for brand-mention detection, by campaign. Finance is excluded from all analyses. Real Estate Sales is excluded from brand similarity (det.\ rate 53.6\% $<$ 70\% threshold) but its lexicon is listed for completeness.}
    \label{tab:brand_list}
    \begin{tabular}{p{3.5cm} p{10.3cm}}
        \toprule
        \textbf{Campaign} & \textbf{Brands tracked (canonical names)} \\
        \midrule
        Telecommunications \newline \textit{(51 brands)} &
            1\&1, ALDI, Alao, Besteabos, CH Mobile, Comparis, Congstar, Coop, Deinabo, Digital Republic, Dschungelkompass, Freshnet, GGA Maur, Galaxus Mobile, Gigamobile, Handyabo-Vergleich, Init7, Jio, Lebara, Leucom, Lidl, Lycamobile, MTEL, Migros, Moneyland, Monzoon, Mucho, Net+, Netplus, Netzwoche, O2, Peoplefone, Post Mobile, Quickline, Sak Digital, Salt, Solnet, Spusu, Sunrise, Swisscom, Swype, TalkTalk, Teleboy, Toppreise, Ubigi, VTX, Vodafone, Wingo, Yallo, gomo, iWay \\[3pt]
        Sporting Goods \newline \textit{(43 brands)} &
            21run, ASICS, Adidas, Altra, Berg-Freunde, Birkenstock, Brooks, Bächli, Decathlon, HOKA, Idealo, Inov-8, Intersport, KURU Footwear, Karhu, Kiprun, La Sportiva, Lauf-bar, Merrell, Mizuno, New Balance, Nike, Norda, OOFOS, Ochsnersport, On, Puma, Reebok, Runnersworld, Running Point, Runningxpert, Salomon, Saucony, Scott, Shop4runners, Skechers, Sportscheck, The North Face, Topo Athletic, Transa, UGG, Under Armour, Vivobarefoot \\[3pt]
        Consumer Electronics \newline \textit{(47 brands)} &
            AMD, ASUS, Acer, Alienware, Alternate, Amazon, Apple, Back Market, Brack, CHUWI, Conrad, Cyberport, Dell, Digitec, Dynabook, ERAZER, Framework, Fujitsu, Fust, Galaxus, Gigabyte, HP, Honor, Huawei, Idealo, Intel, Interdiscount, LG, Lenovo, Logitech, MSI, Mediamarkt, Medion, Microsoft, Microspot, NVIDIA, Notebookcheck, Panasonic, Preisvergleich, Razer, Samsung, Sony, Steg Electronics, Toppreise, Toshiba, VAIO, XMG \\[3pt]
        Real Estate Sales \newline \textit{(excl.; det.\ rate 53.6\%; 32 brands)} &
            AXA, Acheter-Louer, BEKB, Baloise, Beobachter, Blick, CBRE, Comparis, Engel \& Völkers, Finanztip, Helvetia, Homeday, Homegate, ImmoScout24, Immoverkauf24, Immowelt, JLL, Livit, Mobiliar, Neho, Newhome, PostFinance, Privera, Properti, RE/MAX, Raiffeisen, Sotheby's, Swiss Life, UBS, Wincasa, Wüst \& Wüst, ZKB \\[3pt]
        \bottomrule
    \end{tabular}
    \\\small\textit{Note:} Detection is substring-based on lower-cased answer text. Multiple search patterns may map to the same canonical brand name (e.g., \textit{m-budget}, \textit{mbudget} $\to$ Migros; \textit{aldi mobile} $\to$ ALDI). Brand counts per vertical: Real Estate Sales 32, Sporting Goods 43, Consumer Electronics 47, Telecommunications 51.
    \label{tab:brand_lexicon}
\end{table}

\section{Campaign Prompts}
\label{sec:prompts}

The following tables list all eight prompts per campaign as originally issued to the search engines in German, alongside their English translations. Prompts were derived from high-search-volume Swiss SEO keywords using Google's ``People Also Ask'' feature.

\begin{table}[H]
    \centering
    \small
    \caption{Prompts used for the Sporting Goods campaign.}
    \label{tab:prompts_sportartikel}
    \begin{tabular}{p{7.5cm} p{7.5cm}}
        \toprule
        \textbf{German (original)} & \textbf{English (translation)} \\
        \midrule
        Auf was muss man beim Laufschuhkauf achten? & What should you pay attention to when buying running shoes? \\[3pt]
        In welchen Schuhen läuft man wie auf Wolken? & In which shoes do you run as if on clouds? \\[3pt]
        Was ist die 80\%-Regel beim Laufen? & What is the 80\% rule in running? \\[3pt]
        Welche Marke sind gute Laufschuhe? & Which brand makes good running shoes? \\[3pt]
        Wie finde ich die richtigen Laufschuhe für mich? & How do I find the right running shoes for me? \\[3pt]
        Wie teuer muss ein guter Laufschuh sein? & How expensive does a good running shoe have to be? \\[3pt]
        Wie viel kosten sehr gute Laufschuhe? & How much do very good running shoes cost? \\[3pt]
        Wie viel sollte ein Anfänger für Laufschuhe ausgeben? & How much should a beginner spend on running shoes? \\
        \bottomrule
    \end{tabular}
\end{table}

\begin{table}[H]
    \centering
    \small
    \caption{Prompts used for the Consumer Electronics campaign.}
    \label{tab:prompts_elektronik}
    \begin{tabular}{p{7.5cm} p{7.5cm}}
        \toprule
        \textbf{German (original)} & \textbf{English (translation)} \\
        \midrule
        Wann ist die beste Zeit, um einen Laptop zu kaufen? & When is the best time to buy a laptop? \\[3pt]
        Was ist der Unterschied zwischen einem Notebook und einem Laptop? & What is the difference between a notebook and a laptop? \\[3pt]
        Welche Laptops sind zurzeit die besten? & Which laptops are currently the best? \\[3pt]
        Welche Marke ist für Laptops am besten geeignet? & Which brand is best suited for laptops? \\[3pt]
        Welche Marke ist gut bei Laptops? & Which brand is good for laptops? \\[3pt]
        Welcher Laptop wird am meisten gekauft? & Which laptop is purchased most often? \\[3pt]
        Wie viel sollte ein guter Laptop kosten? & How much should a good laptop cost? \\[3pt]
        Worauf sollte man beim Kauf eines Laptops achten? & What should you look for when buying a laptop? \\
        \bottomrule
    \end{tabular}
\end{table}

\begin{table}[H]
    \centering
    \small
    \caption{Prompts used for the Telecommunications campaign.}
    \label{tab:prompts_telekom}
    \begin{tabular}{p{7.5cm} p{7.5cm}}
        \toprule
        \textbf{German (original)} & \textbf{English (translation)} \\
        \midrule
        In der Schweiz: Welche Handy-Abos sind weltweit unlimitiert? & In Switzerland: Which mobile phone plans are unlimited worldwide? \\[3pt]
        In der Schweiz: Welcher Anbieter hat unlimited Datenvolumen? & In Switzerland: Which provider offers unlimited data volume? \\[3pt]
        In der Schweiz: Welcher Internetanbieter ist zurzeit der beste? & In Switzerland: Which internet provider is currently the best? \\[3pt]
        Was ist das schnellste Internet in der Schweiz? & What is the fastest internet in Switzerland? \\[3pt]
        Welche Anbieter in der Schweiz bieten ein Internetabonnement ohne Vertrag an? & Which providers in Switzerland offer an internet subscription without a contract? \\[3pt]
        Welcher Anbieter hat das beste Internet in der Schweiz? & Which provider has the best internet in Switzerland? \\[3pt]
        Welcher Handyanbieter ist der beste in der Schweiz? & Which mobile phone provider is the best in Switzerland? \\[3pt]
        Wie bekomme ich unbegrenzte Daten in der Schweiz? & How do I get unlimited data in Switzerland? \\
        \bottomrule
    \end{tabular}
\end{table}

\begin{table}[H]
    \centering
    \small
    \caption{Prompts used for the Real Estate Sales campaign.}
    \label{tab:prompts_immobilien}
    \begin{tabular}{p{7.5cm} p{7.5cm}}
        \toprule
        \textbf{German (original)} & \textbf{English (translation)} \\
        \midrule
        Ist es aktuell sinnvoll, ein Haus zu verkaufen? & Is it currently advisable to sell a house? \\[3pt]
        Was muss ich alles bezahlen, wenn ich mein Haus verkaufe? & What do I have to pay when I sell my house? \\[3pt]
        Was muss ich beachten, wenn ich mein Haus verkaufen möchte? & What do I need to consider if I want to sell my house? \\[3pt]
        Wie berechne ich den Gewinn einer Immobilie? & How do I calculate the profit from a property? \\[3pt]
        Wie hoch ist der langfristige Kapitalgewinn beim Verkauf einer Immobilie? & How high is the long-term capital gain when selling a property? \\[3pt]
        Wie lange muss ich mein Haus besitzen, um es steuerfrei zu verkaufen? & How long do I have to own my house to sell it tax-free? \\[3pt]
        Wie viel Steuer muss ich zahlen, wenn ich mein Haus verkaufe? & How much tax do I have to pay when I sell my house? \\[3pt]
        Wie viel ist kapitalertragssteuerfrei? & How much is exempt from capital gains tax? \\
        \bottomrule
    \end{tabular}
\end{table}

\section{Source Citation Inequality (Gini)}
\label{sec:gini_appendix}

The heatmap of Gini coefficients by campaign and engine is shown in Figure~\ref{fig:gini_heatmap} in the main text (Section~\ref{sec:temporal}). All campaigns and engines show high Gini values (0.63--0.83), confirming that a small number of domains receive the vast majority of citations. The overall mean Gini is 0.715. Google AI Mode exhibits the highest concentration (0.782) and Perplexity the lowest (0.671). Across campaigns, Telecommunications reaches the highest Gini (0.750) and Sporting Goods the lowest (0.680). Tables~\ref{tab:gini-campaign} and~\ref{tab:gini-searchengine} provide the numerical breakdown by campaign and engine, respectively; the formula and a worked example follow below.

\begin{table}[ht]
\centering
\small
\caption{Mean Gini coefficient by campaign (mean across engines), Jan 24 -- Mar 20, 2026.}
\begin{tabular}{lr}
\toprule
Campaign & Gini \\
\midrule
Consumer Electronics        & 0.713 \\
Real Estate Sales & 0.718 \\
Sporting Goods      & 0.680 \\
Telecommunications           & 0.750 \\
\bottomrule
\end{tabular}
\label{tab:gini-campaign}
\end{table}

\begin{table}[ht]
\centering
\small
\caption{Mean Gini coefficient by search engine (mean across campaigns), Jan 24 -- Mar 20, 2026. \texttt{images.openai.com} CDN artifact excluded (see Section~\ref{sec:limitations}).}
\begin{tabular}{lr}
\toprule
Search Engine    & Gini \\
\midrule
ChatGPT          & 0.684 \\
Gemini           & 0.723 \\
Google AI Mode   & 0.782 \\
Perplexity       & 0.671 \\
\bottomrule
\end{tabular}
\label{tab:gini-searchengine}
\end{table}

\section{Gini Coefficient: Formula, Example, and Implementation}
\label{sec:gini_formula}

\subsection{Formula}

The Gini coefficient \citep{giniMeasurementInequalityIncomes1921} measures inequality in a distribution: $G = 0$ means all domains receive equal citations; $G = 1$ means one domain receives all citations. For a finite set of $n$ non-negative values, sorted in ascending order $y_1 \leq y_2 \leq \cdots \leq y_n$, it is computed as:

\begin{equation}
G = \frac{2 \sum_{i=1}^{n} i \cdot y_i}{n \sum_{i=1}^{n} y_i} - \frac{n+1}{n}
\label{eq:gini-brown}
\end{equation}

where $y_i$ is the citation count of domain $i$ and $i$ its ascending rank. This is the standard rank-weighted computational form, algebraically equivalent to the classical Lorenz-curve definition ($G = 2 \times$ area between the Lorenz curve and the line of equality). It requires only one pass through the sorted data.

\subsection{Worked Example}

Consider five domains with citation counts $[1, 2, 3, 4, 10]$ (already sorted ascending).

\begin{enumerate}
    \item Assign ranks $i = [1, 2, 3, 4, 5]$.
    \item Compute weighted sums:
    \begin{align*}
        \textstyle\sum y_i &= 1 + 2 + 3 + 4 + 10 = 20 \\
        \textstyle\sum i \cdot y_i &= 1{\cdot}1 + 2{\cdot}2 + 3{\cdot}3 + 4{\cdot}4 + 5{\cdot}10 = 80
    \end{align*}
    \item Apply Equation~\eqref{eq:gini-brown}:
    \begin{align*}
        G &= \frac{2 \times 80}{5 \times 20} - \frac{6}{5} = 1.6 - 1.2 = 0.4
    \end{align*}
\end{enumerate}

$G = 0.4$ indicates moderate inequality: the dominant domain (10 citations) accounts for 50\% of all citations while the four others share the remaining 50\% unevenly.

\section{Convergence Analysis: How Many Runs Are Sufficient?}
\label{sec:convergence}

\subsection{Motivation and Method}

The stochastic nature of LLM outputs means that a single query yields only a noisy snapshot of a brand's true visibility. This is analogous to the \emph{pass@k} problem in code generation, where \citet{chenEvaluatingLargeLanguage2021} show that estimating the probability of a correct solution requires multiple independent samples. \citet{liorReliableEvalRecipeStochastic2025} formalise this for general LLM evaluation via the method of moments, deriving the number of repeated runs needed for reliable evaluation. \citet{mizrahiStateWhatArt2024} similarly find that single-prompt LLM evaluation is unreliable and recommend multi-prompt, multi-run designs. We apply the same logic to GEO measurement, asking: how many repeated runs are needed to estimate brand-mention probability reliably?

The minimum-run-count recommendation in the Conclusion (Section~\ref{sec:simul}) is derived empirically from the 10-run simultaneous dataset. The full dataset contains 128 engine--prompt groups (4 engines $\times$ 8 prompts $\times$ 4 campaigns) with all 10 runs available. Brand analysis is restricted to the three qualifying campaigns (96 groups; see Section~\ref{sec:temporal}); source coverage analysis uses all 128 groups. We treat the 10-run mean as the best available proxy for the ``true'' detection probability.

\textbf{Method: subsampling without replacement.} For each group and subsample size $n \in \{1, \ldots, 9\}$, we draw 2{,}000 random subsamples of size $n$ \emph{without} replacement \citep{efronIntroductionBootstrap1994} and record the mean binary detection indicator for each individual brand (1 if that specific brand was detected in the response, 0 otherwise). Rather than collapsing to a campaign-level ``any brand'' indicator, we treat each canonical brand as a separate series. Brands that are never detected across all 10 runs are excluded (their SE is trivially zero and not informative). This yields 1{,}216 per-brand series across the three qualifying campaigns. The standard deviation of the 2{,}000 subsample means is the estimated SE of an $n$-run estimate for that brand; we report the mean SE across all 1{,}259 series. Note that sampling without replacement introduces a finite population correction: at $n = 9$ from $N = 10$ runs, the SE is mechanically smaller than it would be for truly independent additional runs. The reported SE values therefore represent a lower bound; actual SE from fresh data would be somewhat higher at large $n$.

\textbf{Source coverage.} We measure how well an $n$-run union of cited domains approximates the 10-run reference union, using Jaccard similarity between the two. We draw 2{,}000 subsamples per group and report SE of this Jaccard across all 128 groups.

\subsection{Results}

\begin{figure}[ht]
    \centering
    \includegraphics[width=\linewidth]{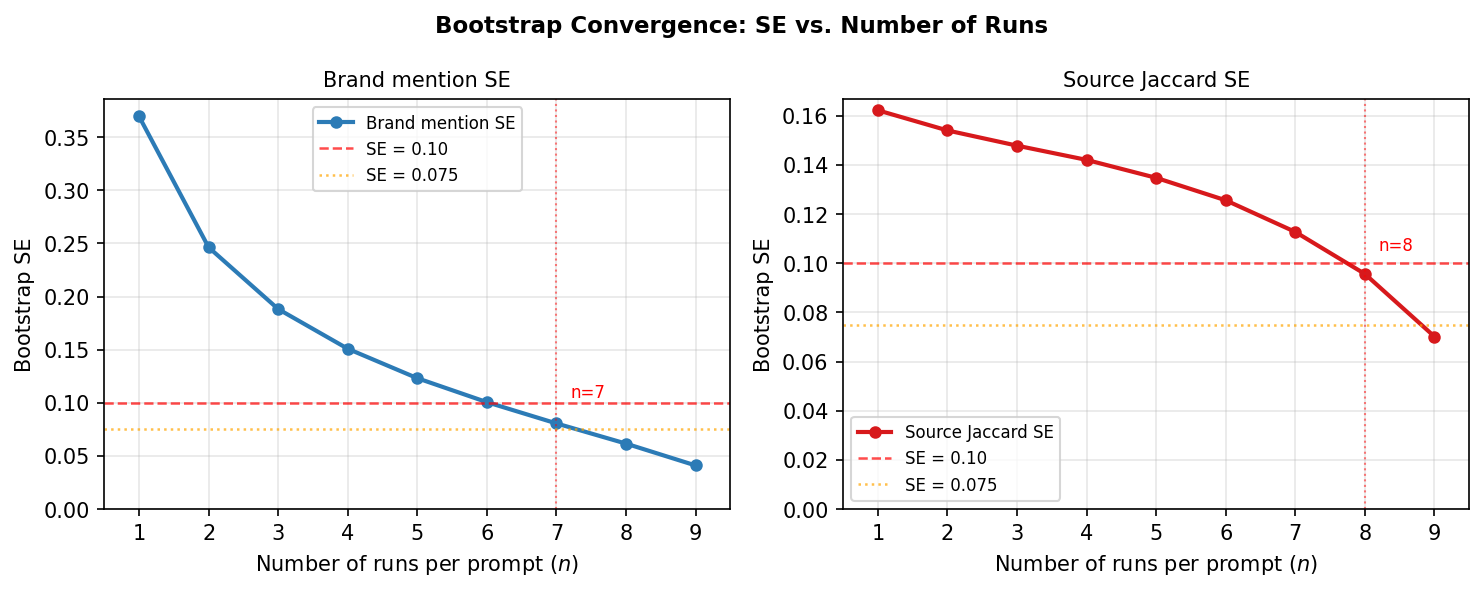}
    \caption{Subsampling standard error of the estimated per-brand detection rate (left) and source-coverage Jaccard (right) as a function of the number of independent runs per prompt. Each point is the mean SE across all 1{,}216 per-brand series (left) or 128 engine--prompt groups (right), using 2{,}000 subsamples without replacement. The red dashed line marks SE~$= 0.10$. Per-brand monitoring reaches SE~$< 0.10$ at $n = 7$ runs; source coverage requires $n = 8$ runs.}
    \label{fig:convergence}
\end{figure}

Figure~\ref{fig:convergence} shows both curves. For per-brand detection rate (left panel), SE falls below 0.10 at $n = 7$ runs (95\% CI $\pm 0.158$) and below 0.08 at $n = 8$ runs (95\% CI $\pm 0.121$). The curve is steep between 1 and 5 runs, and flattens thereafter: moving from 8 to 9 runs only reduces SE from 0.062 to 0.041. For source coverage (right panel), convergence is similar: SE remains above 0.10 until $n = 8$ runs (SE = 0.096, 95\% CI $\pm 0.187$), reflecting the higher stochasticity of which specific URLs are cited.

\begin{table}[ht]
\centering
\small
\caption{Subsampling SE and 95\% confidence interval half-width for per-brand detection rate estimation by number of runs (averaged across 1{,}216 per-brand series).}
\label{tab:convergence}
\begin{tabular}{r cc}
\toprule
\textbf{Runs ($n$)} & \textbf{SE} & \textbf{95\% CI ($\pm$)} \\
\midrule
1 & 0.370 & 0.724 \\
2 & 0.246 & 0.483 \\
3 & 0.188 & 0.369 \\
4 & 0.151 & 0.296 \\
5 & 0.123 & 0.241 \\
6 & 0.101 & 0.197 \\
7 & 0.081 & 0.158 \\
8 & 0.062 & 0.121 \\
9 & 0.041 & 0.081 \\
\bottomrule
\end{tabular}
\\\small\textit{Note:} SE = std of 2{,}000 subsamples without replacement per brand series. ``True'' rate proxied by the 10-run mean for that brand. 96 engine--prompt groups (4 engines $\times$ 8 prompts $\times$ 3 qualifying campaigns) $\times$ multiple brands per group $= $ 1{,}216 per-brand series (brands never detected across all 10 runs excluded). SE at $n = 9$ is subject to finite population correction (FPC) and underestimates the SE of truly independent runs.
\end{table}

\subsection{Interpretation}

A single run (SE = 0.370) is essentially uninformative: a true per-brand detection rate of 50\% could appear anywhere from $-22\%$ to $+122\%$ in a nominal 95\% interval (clipped to [0,1] in practice). At $n = 7$ runs SE drops to 0.081, giving a 95\% CI of $\pm 0.158$---adequate for detecting large differences (e.g., a brand detected in 80\% vs.\ 20\% of runs) but insufficient for fine-grained ranking of brands with similar visibility. At $n = 8$ runs SE falls to 0.062 ($\pm 0.121$), and source coverage reaches comparable precision (SE = 0.096). The per-brand framing is more actionable than a campaign-level ``any brand'' indicator because it surfaces which specific brands are consistently absent and which are reliably cited. These thresholds assume intermediate detection probabilities; for brands that are either always or never cited, fewer runs suffice. This empirical result aligns with the formal framework of \citet{liorReliableEvalRecipeStochastic2025}, who derive minimum sample requirements for reliable LLM evaluation from first principles.

\section{Temporal Convergence: How Long an Observation Window Is Sufficient?}
\label{sec:temporal_convergence}

\subsection{Motivation and Method}

The sustained-observation-window recommendation in the Conclusion is derived empirically from the temporal dataset (Jan 24 -- Mar 20, 2026). Mirroring Appendix~\ref{sec:convergence}, we use a per-brand approach: for each canonical brand in the three qualifying campaigns, we build a daily binary series (1 if that brand was detected in the day's response, 0 otherwise). Brands never detected across the entire observation period are excluded. This yields 1{,}726 per-brand series (3 qualifying campaigns $\times$ 4 engines $\times$ 8 prompts $\times$ multiple brands, filtered to those with $\geq 1$ detection), spanning 40--46 days per series.

We ask: as the rolling window length $d$ increases, how precisely can a practitioner estimate the underlying per-brand detection probability from a $d$-day mean?

For each series and each possible $d$-day consecutive window within its observation period, we compute the mean detection rate. The standard error (SE) across all such window means, averaged over all 1{,}726 series, quantifies estimation uncertainty as a function of window length $d$. This follows the same logic as rolling-window volatility estimation in time-series analysis.

\subsection{Results}

\begin{figure}[ht]
    \centering
    \includegraphics[width=\linewidth]{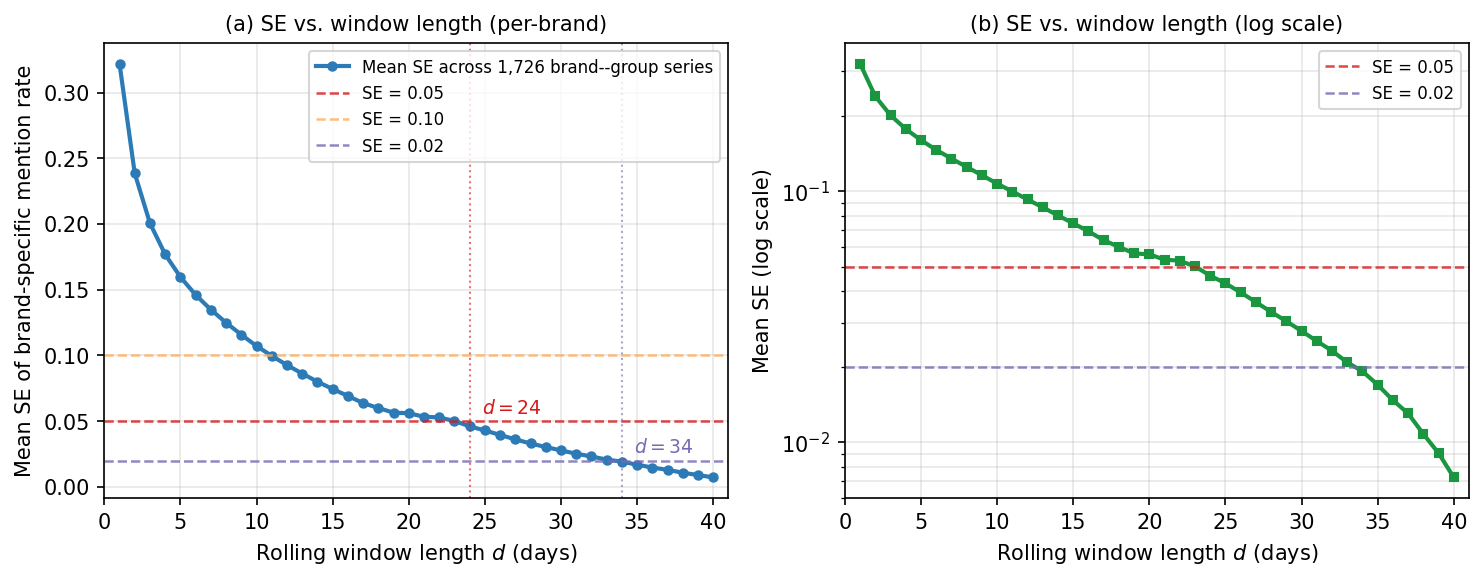}
    \caption{Mean standard error of the $d$-day rolling window per-brand detection rate (averaged across 1{,}726 per-brand series) as a function of window length $d$. Left: linear scale; right: log scale. The red dashed line marks SE~$= 0.05$; the purple dashed line marks SE~$= 0.02$. SE falls below 0.05 at $d = 24$ days and below 0.02 at $d = 34$ days.}
    \label{fig:temporal_convergence}
\end{figure}

Figure~\ref{fig:temporal_convergence} shows the mean SE as a function of window length $d$ (left panel, linear scale; right panel, log scale). Table~\ref{tab:temporal_convergence} lists key thresholds.

\begin{table}[ht]
\centering
\small
\caption{Rolling-window SE and 95\% confidence interval half-width for per-brand detection rate estimation by window length (averaged across 1{,}726 per-brand series).}
\label{tab:temporal_convergence}
\begin{tabular}{r cc}
\toprule
\textbf{Window ($d$ days)} & \textbf{SE} & \textbf{95\% CI ($\pm$)} \\
\midrule
1  & 0.322 & 0.631 \\
2  & 0.238 & 0.467 \\
3  & 0.201 & 0.393 \\
5  & 0.160 & 0.313 \\
7  & 0.135 & 0.264 \\
10 & 0.107 & 0.210 \\
14 & 0.080 & 0.157 \\
21 & 0.053 & 0.105 \\
28 & 0.033 & 0.065 \\
\bottomrule
\end{tabular}
\\\small\textit{Note:} SE computed as the standard deviation of all possible $d$-day window means within each per-brand series. Series: 1{,}726 per-brand indicators across 3 qualifying campaigns $\times$ 4 engines $\times$ 8 prompts (brands never detected excluded). Temporal dataset: Jan 24 -- Mar 20, 2026.
\end{table}

\subsection{Interpretation}

The per-brand convergence is considerably slower than a campaign-level ``any brand'' indicator, reflecting the higher variability of individual brand detection rates. SE falls below 0.10 at $d = 10$ days and below 0.05 at $d = 24$ days (95\% CI $\pm 0.105$ at $d = 21$; $\pm 0.065$ at $d = 28$). A 14-day window still leaves SE at 0.080 ($\pm 0.157$)---sufficient for directional monitoring but not for fine-grained brand comparison. This result must be interpreted in the context of AI search dynamics.

AI search engines undergo regular algorithmic updates and index refreshes that can shift brand inclusion probabilities substantially over days to weeks. Short windows---even when statistically tight at the campaign level---may produce estimates that are unrepresentative of the longer-run visibility level for specific brands. A two-to-four-week rolling window is recommended because it (a) reduces per-brand SE below 0.05--0.08, within practical precision requirements for brand monitoring, and (b) averages over short-lived fluctuations introduced by minor model updates, thereby providing a more durable and actionable estimate of sustained per-brand visibility.

\section{Code and Data Availability}
\label{sec:code_availability}

All analysis code and the datasets used in this study are publicly available at:
\begin{center}
\url{https://github.com/jatlantic/DONT-MEASURE-ONCE-MEASURING-VISIBILITY-IN-AI-SEARCH}
\end{center}
The repository contains the two analysis scripts (\texttt{paper\_analysis\_v9.py}, \texttt{temporal\_brand\_v9.py}), the shared similarity utility (\texttt{similarity\_functions.py}), the simultaneous-run dataset (\texttt{live\_20260321\_042355.jsonl}), the brand lexicon, and the processed temporal data files derived from the Aurora Intelligence export.

\end{document}